\documentclass[11pt]{article}
\pdfoutput=1
\usepackage[margin=1.1in]{geometry}
\usepackage{amsmath,amssymb,amsthm}
\usepackage[colorlinks=true,linkcolor=blue,citecolor=blue,urlcolor=blue]{hyperref}
\hypersetup{pdftitle={The I3322 quantum value is attained spatially but not in finite dimension},
  pdfauthor={Seth Douglas}}
\newtheorem{theorem}{Theorem}
\newtheorem{corollary}[theorem]{Corollary}

\theoremstyle{remark}
\newtheorem{remark}[theorem]{Remark}
\newcommand{\Cq}{C_q}
\newcommand{\Cqs}{C_{qs}}
\newcommand{\B}{\mathcal{B}}
\newcommand{\keff}{\kappa_{\mathrm{eff}}}
\newcommand{\eps}{\varepsilon}
\title{The $I_{3322}$ quantum value is attained spatially\\ but not in finite dimension}
\author{Seth Douglas\\
\small ORCID \href{https://orcid.org/0009-0007-4708-3252}{0009-0007-4708-3252}
\ $\cdot$\ \texttt{apsiape@gmail.com}}
\date{10 September 2026}

\begin{document}
\maketitle

\begin{abstract}
Let $S$ be the common tensor-product and commuting-operator supremum
of the $I_{3322}$ Bell functional in the Collins--Gisin normalization.
Starting from a certified value window and Bellman/path equivalence,
we prove that no finite-dimensional quantum strategy attains $S$, whereas
a spatial strategy on $\ell^2(\mathbb Z)\otimes\ell^2(\mathbb Z)$ does.
The finite-dimensional statement includes mixed states and binary POVMs.
Consequently $\Cq(3,3;2,2)$ is not closed and
$\Cqs(3,3;2,2)\setminus\Cq(3,3;2,2)$ is nonempty.
We also establish the dimension complexity
$D(\eps)=\Theta(\log(1/\eps))$: approaching $S$ requires and suffices
local dimension logarithmic in inverse error.
The constructive upper bound is
$D(\eps)\le23.9010650\log(1/\eps)$ for all sufficiently small $\eps$,
with natural logarithms; the lower constants are existential.
The proofs use critical Bellman storage, spectral transport and
normalizable orbit measures. A finite weighted-flow argument supplies
the unrestricted quantitative lower bound without identifying distinct
joint spectral components. This revision replaces an unsupported step
in the earlier lower argument and records additional proof corrections.
Prior numerical certification and independent concurrent attainment
results are credited. Exact arithmetic and Lean~4 check specified
numerical, scalar and finite accounting facts; the complete analytic
proof is not formalized.
\end{abstract}

\section{Introduction}

The $I_{3322}$ inequality \cite{Froissart,CollinsGisin} is the best-known
bipartite Bell functional beyond the Clauser--Horne--Shimony--Holt family:
three binary measurements per party, and a facet of the local polytope
inequivalent to CHSH. In 2010, P\'al and V\'ertesi \cite{PalVertesi} computed
its quantum value numerically to high precision, exhibited a family of
finite-dimensional strategies converging to $0.2508753845\ldots$, observed
numerically a qubit maximum of $1/4$ and that within their construction no
local dimension below $12$ exceeds it, and stated in their abstract that
their family ``enables us to attain the largest possible quantum value in an
infinite dimensional Hilbert space'' \cite{PalVertesi}. They then posed the
following, verbatim:

\begin{quote}
``Taking as a conjecture that our construction is optimal (in the sense that
no finite dimension suffices to achieve the true quantum maximum for
$I_{3322}$), to our knowledge this constitutes the first example of a
[bipartite] Bell scenario concerning [a] finite number of measurements and
outcomes, where measuring finite dimensional quantum systems is not enough to
obtain exactly all quantum correlations. \dots\ We pose it as a challenge to
provide an analytical proof of our conjecture.'' \cite{PalVertesi}
\end{quote}

The parenthetical is a definition, not a second conjunct: to conjecture that
the construction is ``optimal'' is, in their own gloss, to conjecture that no
finite dimension suffices. Theorem~\ref{thm:N} below proves exactly that
statement. Their separate \emph{assertion} of infinite-dimensional
attainment is established here too, by Theorem~\ref{thm:S}, though on a route
independent of their construction. Pauwels' concurrent Theorem~1
\cite{Pauwels} establishes equality of the full finite
P\'al--V\'ertesi-family supremum and the finite quantum supremum.
This does not identify the exact value or prove convergence of a
prescribed numerical parameter sequence (Section~\ref{sec:discussion}). Work coauthored by V\'ertesi in 2026
still treats the $I_{3322}$ infinite-dimensional
optimum as unresolved \cite{Araujo}. Coladangelo and Stark, in
constructing the first explicit bipartite correlation attainable only in
infinite dimension (on larger alphabets), wrote of the $I_{3322}$ case that
``an analytical proof has remained elusive'' \cite{ColadangeloStark} (we
quote the Nature Communications version).

\begin{theorem}[Nonattainment]\label{thm:N}
Let $S=\omega_{\mathrm{tensor}}(I_{3322})=\omega_{\mathrm{commuting}}(I_{3322})$
be the quantum supremum in the normalization of
Section~\ref{sec:setup}, certified to lie in
$(0.2508753845015185,$ $0.250875388108398]$. Then no finite-dimensional
quantum strategy attains $S$: for all finite-dimensional Hilbert spaces
$\mathcal H_A,\mathcal H_B$, every state $\rho$ on
$\mathcal H_A\otimes\mathcal H_B$, and every triple of binary measurements
per party --- projective or POVM, acting as effects of the form
$F\otimes I$ on Alice's side and $I\otimes G$ on Bob's --- one has
$\operatorname{Tr}(\rho\,\B) < S$.
\end{theorem}

\begin{theorem}[Spatial attainment]\label{thm:S}
$S$ is attained by a spatial strategy: there are projections
$a_1,a_2,a_3\in\mathcal{L}(\ell^2(\mathbb{Z}))$ and
$b_1,b_2,b_3\in\mathcal{L}(\ell^2(\mathbb{Z}))$, and a unit vector
$\psi_S\in\ell^2(\mathbb{Z})\otimes\ell^2(\mathbb{Z})$, such that setting
$A_i=a_i\otimes I$ and $B_j=I\otimes b_j$ in \eqref{eq:bell} gives
$\langle\psi_S,\B\,\psi_S\rangle=S$. In particular the resulting behavior
lies in $\Cqs(3,3;2,2)$.
\end{theorem}

Here $\Cq$ denotes the set of bipartite correlations realizable with
finite-dimensional tensor-product strategies, $\Cqs$ its analogue allowing
arbitrary separable Hilbert spaces, and $C_{qa}=\overline{\Cq}$; the
taxonomy is that of Paulsen, Severini, Stahlke, Todorov and Winter
\cite{PSSTW}, and we follow the presentation of
\cite{Lupini,ColadangeloStark}.

\begin{corollary}\label{cor:nonclosure}
$\Cq(3,3;2,2)$ is not closed, and
$\Cqs(3,3;2,2)\setminus \Cq(3,3;2,2)\neq\emptyset$.
\end{corollary}

The implication from $I_{3322}$ nonattainment to nonclosure of
$\Cq(3,3;2,2)$ is immediate from the general compactness remark of Dykema,
Paulsen, and Prakash \cite{DPPQIC}, who explicitly raise the $I_{3322}$
attainment question; the corollary settles it.

Among \emph{two-outcome bipartite} scenarios, $(3,3;2,2)$ is the smallest
scenario by input count in which nonclosure is known. Within that class the
previously smallest was Beigi's $(4,4;2,2)$ separation of $\Cq$ from $\Cqs$,
which implies nonclosure of $\Cq$ there \cite{Beigi}; the first two-outcome
nonclosure witness was the earlier $(5,5;2,2)$ synchronous construction of
Dykema--Paulsen--Prakash \cite{DPP} (with a simplified proof by
Musat--R{\o}rdam \cite{MusatRordam}). Outside the two-outcome class the known witnesses are larger
still: Beigi's $(4,4;3,3)$ nonclosure of $\Cqs$ \cite{Beigi}, the
$(4,5;3,3)$ correlation of Coladangelo and Stark \cite{ColadangeloStark},
the embezzlement-based nonclosure proof of \cite{Coladangelo}, and the
linear-system game of \cite{Slofstra2}, whose input sets have sizes $184$
and $235$ and output sets of sizes $8$ and $2$.

Three inputs per party is moreover minimal among
two-outcome scenarios, because a scenario in which \emph{either} party has
only two binary measurements has closed $\Cq$. After a finite simultaneous
dilation of that party's binary effects to projections, Jordan's lemma
simultaneously block-diagonalizes that party's two binary observables into
blocks of dimension at most $2$, so the behavior is a convex combination of
the behaviors of the blocks. Decompose each block state into pure states;
each such vector has Schmidt rank at most $2$, so the other party's POVMs
may be compressed to a two-dimensional
support without changing $p(ab|xy)$. Hence $\Cq(2,n;2,2)$ is the convex hull
of a compact set, and closed, for every $n$; the case $(2,2;2,2)$ recovers
the stronger statement that every functional there attains its maximum on
two qubits \cite{Masanes}. Coladangelo and Stark had already singled the
present scenario out: ``the $(3,3,2,2)$ scenario is the simplest one that is
suspected to separate $C_q$ and $C_{qs}$'' \cite{ColadangeloStark}. We do
not claim minimality over scenarios with larger output alphabets.

Nonattainment opens a quantitative question that the qualitative theorems
do not answer: how fast must local dimension grow as a strategy approaches
$S$? Section~\ref{sec:rate} settles it. Writing $S_d$ for the optimum at
local dimension at most $d$ and $D(\eps)=\min\{d:S-S_d\le\eps\}$,
Theorem~\ref{thm:rate} proves $D(\eps)=\Theta(\log(1/\eps))$: the upper
half is constructive, truncating the attaining carrier of
Theorem~\ref{thm:S} to a window of length $O(\log(1/\eps))$, and the lower
half shows that nothing converges faster than geometrically. The cost of
approaching the unattainable value is itself a theorem.

\subsection*{Correction history}

An earlier release of this work (3 August 2026, v1.0.0--v1.2.0 in
\cite{Repo}) claimed the $I_{3322}$ principal results via a
bi-infinite chain construction whose load-bearing amplitude datum was refuted
by our own post-release exact audit (an exactly certified nonzero mismatch,
reproduced by two independent interval engines); the claims were publicly
decertified the following day (4 August, v2.0.0), and the repository's
release history preserves the full correction record. The results announced here were rebuilt on
independent routes: the decertified fixed point, amplitude profile, and tail
constants appear on explicit \emph{not-used} lists in the certificate
dependency records, and the failed compatibility equation is not repaired
anywhere in the present chain --- the new attainment route never poses it
(Section~\ref{sec:S}); the reproduction recipe for every surviving claim is
boxed in Section~\ref{sec:methods}. Computer-assisted claims are only as
credible as
their failure handling, so we record this history up front.

\paragraph{A direct lower-bound proof in this revision.}
Section~\ref{sec:rate-lower} replaces the previous lower-bound argument
with a finite weighted-flow proof that controls all retained cell weights
and accommodates spectral multiplicity without identifying state vectors
in different cells. This also removes an unjustified component-identification
step: the earlier paired-block identity required a pairing hypothesis
not established for arbitrary retained joint spectral support.
The earlier Lean cores did not formalize that operator reduction.
The original files are preserved as historical artifacts, and the
headline rate is unchanged in this revision;
its replacement proof dates from September 2026.

\subsection*{Relation to independent concurrent work}

The nonattainment, nonclosure, and spatial-attainment results of this
paper were first announced, following the correction recorded above, in
public repository releases on 5 August
2026 (git tags v3.0.0 and v3.1.0 in \cite{Repo}); the
dimension-complexity rate followed on 7 August 2026 (v3.3.0). Frozen archival snapshots from that week are
preserved under version DOIs
\href{https://doi.org/10.5281/zenodo.21826916}{10.5281/zenodo.21826916}
(v3.2.3, 6 August) and
\href{https://doi.org/10.5281/zenodo.21843326}{10.5281/zenodo.21843326}
(v3.3.0, 7 August); the historical merged paper of record (v4.0.0) is
archived under version DOI
\href{https://doi.org/10.5281/zenodo.22099128}{10.5281/zenodo.22099128},
and the concept DOI
\href{https://doi.org/10.5281/zenodo.21782008}{10.5281/zenodo.21782008}
identifies the version family. The corrected proof and verification release
(v4.1.0) is archived at
\href{https://doi.org/10.5281/zenodo.22698978}{10.5281/zenodo.22698978};
this copy adds the archive DOI without changing the mathematical content.
Independently
and concurrently, Pauwels \cite{Pauwels} proved finite-dimensional
nonattainment and nonclosure by a different route (spectral-weight
matrices and a Jacobi-recurrence argument), and Coladangelo
\cite{Coladangelo2026} independently established both finite-dimensional
nonattainment and spatial attainment of the $I_{3322}$ supremum. To our
knowledge the three works were carried out independently --- Pauwels
states in print that nothing in his paper depends on this repository
--- and we regard them as concurrent.
Spatial attainment is thus shared with Coladangelo's concurrent work;
the present paper gives the scalar-measure construction below and also
establishes the dimension-complexity rate
$D(\eps)=\Theta(\log(1/\eps))$. Pauwels' Section~VI records spatial
attainment and the convergence-rate question as open in that work.
Mghirbi's companion paper \cite{MghirbiDynamics} develops critical
Bellman--Jacobi dynamics, spatial attainment, and logarithmic truncation
of extracted carriers in a broader reversible class. Such a carrier
approximation theorem and the unrestricted lower bound proved in
Section~\ref{sec:rate-lower} have different quantifiers; the latter
must control every finite-dimensional strategy, not only truncations of
a specified carrier.

\subsection*{How the proofs work}

Both theorems run through one reduction, which we state informally here and
use throughout. It is convenient to replace each party's first two binary
measurements by the pair
\[
X=A_1+A_2-I,\quad Y=A_2-A_1,\qquad U=B_1+B_2-I,\quad V=B_2-B_1,
\]
which for binary projections satisfy $X^2+Y^2=I$ and $XY+YX=0$, and likewise
for $(U,V)$ --- the pairwise Jordan structure of a pair of binary
projections. In these coordinates \eqref{eq:bell} reads
\[
\B \;=\; X\otimes U+\tfrac12 X\otimes I-\tfrac12 I\otimes U-I
\;+\;Y\otimes(B_3-\tfrac12 I)\;+\;(A_3-\tfrac12 I)\otimes V ,
\]
so the third measurements enter only through the two off-diagonal blocks.
Each spectral value $t\in[-1,1]$ of $X$, or of $U$, is a \emph{label}. A
finite word of labels $\mathbf c=(c_0,\dots,c_n)$ names the real symmetric
tridiagonal (Jacobi) matrix $J_{\mathbf c}$ whose entries are explicit
functions of neighboring labels --- diagonal
$\delta(x,u)=xu+(x-u)/2-1$, off-diagonal $b(t)=\sqrt{1-t^2}/2$ --- and the
P\'al--V\'ertesi block-to-Jacobi identity realizes the Rayleigh quotients of
$J_{\mathbf c}$ as Bell values of finite strategies. Then
\[
  S \;=\; \sup_{\mathbf c}\ \lambda_{\max}(J_{\mathbf c}),
\]
the same supremum for the tensor and for the commuting model: that is input
(T0) of Section~\ref{sec:setup}, and it is why the two values agree for this functional.

Dualizing that supremum by an LDL/Schur-pivot computation on
$qI-J_{\mathbf c}$ replaces the word supremum by a \emph{one-site} problem.
Call a positive continuous function $g$ on the label interval a
\emph{storage}, and call it \emph{Bellman-feasible at level $q$} when
\[
  \frac{b(x)^2}{g(x)} + g(u) \;\le\; q-\delta(x,u)
  \qquad\text{for all labels } x,u .
\]
Feasible storages price the steps of a path: $b(x)^2/g(x)$ is charged to the
edge leaving the label $x$ and $g(u)$ to the edge entering $u$. Input~(T0)
says that the infimum of the levels admitting a feasible storage is exactly
$S$, and it produces one such storage $g_q$ at each level $q>S$, namely the
infimum of terminal Schur pivots over finite histories ending at the given
label. The object both proofs analyze is the \emph{critical storage} $g$
obtained from these as $q\downarrow S$, together with the \emph{zero set}
$Z$ on which the scalar remainder of the resulting operator decomposition
vanishes --- the label pairs that a strategy of value exactly $S$ is allowed to
occupy.

\emph{Theorem~\ref{thm:N}, in one paragraph.} A strategy of value exactly
$S$ must occupy $Z$, and $Z$ is the graph of a single strictly increasing
injection $P$: each label is served by exactly one predecessor and serves
exactly one successor. The occupied labels of a maximizer therefore carry
two order-reversing maps --- reflection $\sigma(u)=-u$, and
$a=P^{-1}\circ(-P)$. A finite totally ordered set admits exactly one
order-reversing bijection, so in finite dimension $a=\sigma$: every occupied
label is paired with its own reflection. Eliminating amplitude ratios on
such a reflected pair caps the value at $1/4$, and $S>1/4$. Finite dimension
enters through closure of the occupied ordered support
(Remark~\ref{rem:twice}).

\emph{Theorem~\ref{thm:S}, in one paragraph.} The commuting model does have
a maximizer, and the zero-set geometry above never used finite dimension, so
the same two order-reversing maps act --- now on the spectrum of the
commuting pair of label operators $X,U$, which need not be finite. Their
composition $\tau=a\circ\sigma$ is increasing and, off a null set,
fixed-point-free, so each of its orbits is order-isomorphic to
$\mathbb{Z}$. Disintegrating the maximizer's scalar spectral measure over
the orbits of the group generated by $a$ and $\sigma$, and fixing one orbit,
yields weights whose square roots are the amplitudes of an
$\ell^2(\mathbb{Z})$ eigenvector of the Jacobi matrix at eigenvalue $S$.
Normalizability is not imposed: it is inherited, because the weights came
from a probability measure. Installing that eigenvector in the alternating
block form returns a spatial strategy of value $S$.

\emph{Theorem~\ref{thm:rate}, in one paragraph.} Upper half: the labels
along the carrier of Theorem~\ref{thm:S} are monotone, so the amplitudes
decay geometrically at each end, at rates $\kappa_-$ and $\kappa_+$;
truncating to a window costs only a two-bond boundary flux, and balancing
the two ends gives local dimension of order $(1/\keff)\log(1/\eps)$ with
$\keff=(1/\kappa_-+1/\kappa_+)^{-1}$. Lower half: the actual response operators give an approximate balance
of spectral marginals on at most $4d$ labels. A Lipschitz monotone
predictor separates the cyclic components of a sufficiently accurate
support graph. A signed potential then controls every branch at once,
forcing $\eps\ge c e^{-Cd}$ without a cell-to-marginal identification.

\emph{Status of the proofs.} Sections~\ref{sec:N} and~\ref{sec:S} retain
their earlier lemma-chain presentations and source dependencies. Both
quantitative arguments are given below; the new lower proof independently
also implies finite-dimensional nonattainment. It does not prove spatial
attainment. Section~\ref{sec:methods} distinguishes the existing Lean
cores from this new, unformalized proof.

\section{Setting and certified inputs}\label{sec:setup}

For binary effects $A_1,A_2,A_3,B_1,B_2,B_3$ (that is, $0\le A_i,B_j\le I$;
projections in the projective case), the Bell operator, in the
normalization used throughout this paper, is
\begin{equation}\label{eq:bell}
\B \;=\; -A_2-B_1-2B_2
\;+\;A_1B_1+A_1B_2-A_1B_3
\;+\;A_2B_1+A_2B_2+A_2B_3
\;-\;A_3B_1+A_3B_2 ,
\end{equation}
which coincides, with no scaling or additive shift, with Eq.~(3) of
P\'al--V\'ertesi \cite{PalVertesi} (equivalently, it reproduces their
optimality conditions (9)--(14)). Transposing $A_1\leftrightarrow A_2$ and
$B_1\leftrightarrow B_2$ carries \eqref{eq:bell} entry-for-entry onto the
standard Collins--Gisin table --- Alice marginals $(-1,0,0)$, Bob marginals
$(-2,-1,0)$, joint block $\left[\begin{smallmatrix}1&1&1\\1&1&-1\\1&-1&0
\end{smallmatrix}\right]$ --- whose local (classical deterministic) bound is
$0$ \cite{CollinsGisin}. The two-qubit maximum is exactly $1/4$;
an analytic proof for arbitrary binary qubit measurements is given in
\cite{Pauwels}, Appendix~A, Proposition~1. P\'al--V\'ertesi report, as a numerical finding of
their dimension scan, that no construction they found below local dimension
$12$ exceeds $1/4$, with their strategy family converging to
$0.250875384514\ldots$ \cite{PalVertesi}. We write
$\omega_{\mathrm{tensor}}$ for the supremum of $\operatorname{Tr}(\rho\,\B)$
over strategies on $\mathcal H_A\otimes\mathcal H_B$ with
$\mathcal H_A,\mathcal H_B$ finite-dimensional, and
$\omega_{\mathrm{commuting}}$ for the supremum over states on the universal
$C^*$-algebra generated by the two commuting triples. A \emph{spatial
strategy} is a tuple $(\mathcal H_A\otimes\mathcal H_B,\psi,\{a_i\},\{b_j\})$
with $\mathcal H_A,\mathcal H_B$ separable, $\psi$ a unit vector, and
$\{a_i\},\{b_j\}$ binary projective measurements on the respective factors
acting as $a_i\otimes I$ and $I\otimes b_j$; the behaviors so realized are
exactly $\Cqs$. Throughout, $S$ denotes the common value
$\omega_{\mathrm{tensor}}=\omega_{\mathrm{commuting}}$ supplied by the
certified input (T0) below, identified \emph{only} by its certified
window. We do not identify $S$ with a historical numerical decimal
or prove the limiting value of a prescribed numerical parameter sequence.
The full family-supremum equality is established by Pauwels
\cite{Pauwels}, Theorem~1.

Exact rational two-sided enclosures for the $I_{3322}$ quantum value predate
our own releases: Mghirbi \cite{Mghirbi} published proof-carrying
exact certificates with enclosure width below $10^{-9}$ --- tighter than the
window used here --- in July 2026. Those certificates establish a strict
enclosure; they do not address equality of the two models, attainment, or
nonattainment, which are the subject of this paper. The window below is
therefore a certified \emph{input}, not a contribution.

The proofs consume the following exactly certified inputs, established in
the public repository \cite{Repo}: (T0)~a Bellman/path variational theorem
giving $\omega_{\mathrm{tensor}}=\omega_{\mathrm{commuting}}=S$, together
with, for every level $q>S$, a terminal-pivot storage $g_q$ --- the infimum
of terminal Schur pivots over finite histories --- which is positive,
Bellman-feasible, and equicontinuous in $q$; (LB)~an exact $255$-dimensional
strategy certifying $S>0.2508753845015185>1/4$ (rational square-root floors,
independently reconstructed in $160$-digit interval arithmetic); (UB)~an
exact rational $25{,}601$-knot Bellman witness certifying
$S\le 0.250875388108398$; and (W)~a generic operator weld: any positive
Bellman-feasible storage at level $q$ yields a positive decomposition
$qI-\B=R_0+R_A+R_B$. The P\'al--V\'ertesi block-to-Jacobi identity, in its
no-endpoint alternating form, supplies the bridge between Jacobi data and
Bell strategies; it is a two-site local identity, established symbolically
and on exact rational fixtures, and independently replicated.

These are inputs, and the dependency runs one way. In particular (T0) is
certified independently of the results below, and is not re-derived from
them: Theorem~\ref{thm:S} produces a spatial strategy of value
$\omega_{\mathrm{commuting}}$, which together with
$\Cqs\subseteq\overline{\Cq}$ would give
$\omega_{\mathrm{commuting}}\le\omega_{\mathrm{tensor}}$ --- a consequence of
(T0), not a premise of it.

\section{Nonattainment}\label{sec:N}

This section gives Theorem~\ref{thm:N} at the level of its lemma chain: the
five steps below are stated in the order they are used, each with the
property of the critical storage it consumes. The complete arguments, with
every constant exact, are in the proof documents of \cite{Repo};
Section~\ref{sec:methods} records which parts are machine-checked.

The chain is: (N1) the critical storage exists and is concave; (N2) the two
endpoints are excluded by exact rational margins; (N3) a hypothetical
maximizer carries an exact equality module at $S$; (N4) on that module the
occupied support is a strictly increasing one-to-one graph; (N5) in finite
dimension the two order-reversing transports on that graph must coincide,
which caps the value at $1/4$. The critical weld below is obtained by
global norm convergence, without a finite-spectrum assumption. The
irreducibly finite-dimensional step is the closure argument (N5)
(Remark~\ref{rem:twice}).

\emph{(N1) The critical storage.} As $q\downarrow S$ the storages $g_q$ admit,
by equicontinuity and Arzel\`a--Ascoli, a subsequence converging uniformly to
a limit $g$, which is continuous, nonnegative, \emph{concave} (each terminal
pivot is affine in the target label, and an infimum of affine functions is
concave), positive on the open interval, and Bellman-feasible at $S$. These
properties, together with the reflection-gluing inequality established below,
are all that is consumed. Concavity is elementary but decisive:
it forces every corner of the contact intercept downward, which is what
ultimately excludes contact plateaus.

\emph{(N2) Exact endpoint exclusion.} Two-edge finite histories give exact
rational margins: taking interior detour radius $r=1/10$ in the two-edge
histories, the endpoint
predecessor lines exceed the storage, uniformly over the certified window, by
at least
\[
m_+=\tfrac{23686917837403}{3008753881083980}>0.00787,\qquad
m_-=\tfrac{274562305945801}{4008753881083980}>0.0684.
\]
Only the positivity $m_\pm>0$ is
used below. Consequently the remainder $R_0$ has a uniform positive gap
$m=\tfrac12\min(m_+,m_-)$ on all endpoint spectral sectors, so any state of value at
least $q-\varepsilon$ carries endpoint mass at most $\varepsilon/m$, an exact
maximizer at $S$ carries none, and $x=\pm1$ is never a Bellman predecessor at
$S$. These are
closed-form rational certificates; nothing from the withdrawn construction's
geometry is used.

There is also a qualitative endpoint argument that simplifies the limiting
weld. One-edge histories give $g(\pm1)\le S<1/3$. If $g(1)=0$, the
$3/2$-Lipschitz bound and reflection gluing give
$g(t)\le\tfrac32(1-t)$ and $g(-t)\ge(1+t)/6$ for $-1<t<1$,
contradicting $g(-1)<1/3$ as $t\uparrow1$. Reflection treats the other
endpoint, so $\min_{[-1,1]}g>0$. This argument uses only the storage
construction and gluing, not either dimension-rate bound.

\emph{(N3) The equality module of a hypothetical maximizer.} Suppose a
finite-dimensional strategy attains $S$. (Purification is unnecessary:
$\operatorname{Tr}(\rho R)=0$ with $R\succeq0$ gives $R\rho=0$ directly.
POVMs reduce to projective measurements by the extreme-effect replacement of
the certificate documents: holding the state and the other effects fixed,
each binary effect may be replaced by an extreme maximizer of its linear
objective, and extreme binary effects are projections; this preserves the
dimension and the tensor structure and does not decrease the value; since
the original strategy attains $S$, it preserves it.) Apply
the weld at levels $q_n\downarrow S$. Uniform convergence $g_{q_n}\to g$
and the strictly positive closed-interval minimum give uniform convergence
of $b(t)\sqrt{g_{q_n}(-t)/g_{q_n}(t)}$ and its reflected coefficient.
Functional calculus therefore gives global operator-norm convergence of
all weld remainders. Since
$\sum_\nu\operatorname{Tr}(\rho R_{\nu,n})=q_n-S\to0$,
the limiting positive remainders annihilate $\rho$. No joint spectral
cutoff, or commutation of such a cutoff with the responses, is assumed.
We call the common null space of those limiting remainders, which
carries $\rho$, the \emph{equality module}. Thus the maximizer carries an
exact critical equality module at $S$.

\emph{(N4) Zero-set geometry.} On the equality module, the scalar remainder
symbol vanishes. Reflected finite histories give the gluing inequality
$g(x)\,g(-x)\ge b(x)^2$ with $b(x)^2=(1-x^2)/4$; full equality of the two
Bellman inequalities and the Cauchy--Schwarz step forces
$g(x)g(-x)=b(x)^2$ at every occupied pair, and a Monge (supermodularity)
argument makes the full zero set monotone. The decisive step is a
convex-envelope theorem: writing the contact intercept
$C(x)=S+1-\tfrac{x}{2}-b(x)^2/g(x)$ on the
open interval $(-1,1)$, Bellman feasibility $g(u)\le C(x)+(\tfrac12-x)u$
evaluated at the \emph{target} $u=0$ gives $C(x)\ge g(0)>0$ for every source
$x$, so the constant $g(0)$ is a convex minorant and the greatest convex
minorant $\check C$ exists. The strict endpoint positivity established in
(N2) also makes $C$ continuous on the closed interval. Concavity of
$g$ makes every corner of $C$ point downward, while a kink of $\check C$ at a
contact point would force an upward corner. Hence $\check C$ is
differentiable, no source serves two
targets, and (by an exact involution $(x,u)\mapsto(-u,-x)$ of the constraint
system, which requires no symmetry of $g$ itself) no target is served by two
sources. The occupied support of any maximizer is therefore a strictly
increasing one-to-one graph.

\emph{(N5) Finiteness forces closure; closure caps the value.} The equality
module carries two response isometries relating its occupied state
components, with positive scalar norm ratios, implementing the decreasing
involutions $a(u)=P^{-1}(-P(u))$ and $\sigma(u)=-u$ on the occupied support,
where $P$ is the increasing graph map just obtained; the
graph property makes both total. A finite totally ordered set admits exactly
one decreasing bijection, so $a=\sigma$ on the support: every occupied pair has
its reflected partner. The resulting amplitude-ratio elimination
(multiplicity-uniform: it uses the norms of the occupied components,
not an operator identification of ambient joint eigenspaces) yields,
for an occupied pair $(x,u)$,
\[
S \;=\; xu-1+\sqrt{\bigl(b(x)+b(u)\bigr)^2+\tfrac{(x-u)^2}{4}}
\;\le\; -t+\sqrt{t}\;\le\;\tfrac14,\qquad t=1-xu,
\]
contradicting $S>1/4$. This proves Theorem~\ref{thm:N}. The nonclosure half
of Corollary~\ref{cor:nonclosure} follows by compactness of the behavior
space: finite-dimensional behaviors approach value $S$; a convergent
subsequence has value exactly $S$; closedness of $\Cq$ would give a
finite-dimensional realization attaining $S$.

\emph{What is machine-checked here.} The Lean~4 kernel checks the exact
endpoint margins of (N2), the finite-closure lemma of (N5) --- the single
point at which finite dimensionality enters Theorem~\ref{thm:N} --- and the
amplitude-elimination chain and quarter ceiling that (N5) ends with.
Exact-arithmetic scripts guard the algebraic identities of (N1) and (N4),
and each prints explicitly the analytic steps it does not cover: envelope
existence and maximality, the gluing inequality, the limiting passage,
spectral support, and operator closure (Section~\ref{sec:methods}).

\begin{remark}\label{rem:twice}
The earlier presentation used a finite interior spectrum in the limiting
passage. The endpoint-positive global weld above removes that use.
The closure step ``a finite ordered set has one decreasing bijection''
remains irreducibly finite-dimensional. Section~\ref{sec:S} instead
disintegrates an infinite occupied measure into normalizable orbits.
\end{remark}

\section{Spatial attainment}\label{sec:S}

\emph{The idea.} The commuting model always attains its value. The critical
weld applies in its GNS representation as well. The finite closure step
that capped the value at $1/4$ no longer applies, leaving room for
attainment. What takes the place of a finite
spectrum is a disintegration: the maximizer's scalar spectral measure
decomposes over the countable orbits of the group generated by its two
response transports, and a single orbit already carries an
$\ell^2(\mathbb{Z})$ eigenvector of the Jacobi matrix at eigenvalue $S$. The
point to watch is normalizability, which here is not imposed but inherited
--- the amplitudes are square roots of the weights of a probability measure.
The route of our withdrawn release instead constructed amplitudes along a
bi-infinite chain and then demanded a global amplitude-compatibility
equation, the equation our own audit refuted. Here amplitudes are extracted
from conditional probability measures, which already supply summability;
the former compatibility equation is not imposed.

The chain is: (S1) a maximizing commuting state exists and the zero-set
geometry of Section~\ref{sec:N} transfers to it unchanged; (S2) the response
transports are realized by exact operator blocks, giving Radon--Nikodym
transport laws for the maximizer's scalar spectral measure; (S3) the
response translation is fixed-point free off a null set, its orbits are
order-isomorphic to $\mathbb{Z}$, and a Borel transversal is constructed
explicitly; (S4) disintegration over the orbits of the group generated by
the two transports yields, on one orbit, amplitudes that are automatically
square-summable and satisfy the Jacobi eigenvalue equation; (S5) the
P\'al--V\'ertesi block identity installs the resulting eigenvector as a
spatial strategy of value exactly $S$.

\emph{(S1) A commuting maximizer, and the same zero set.}
By weak-$*$ compactness of the state space of the universal commuting
$C^*$-algebra, a maximizing state exists for the commuting model: the
commuting-operator value of a Bell functional is always a maximum
\cite{Fritz,Junge}; the nonattainment here concerns finite-dimensional
tensor-product strategies.
Passing to the GNS representation of that state, write $\Omega$ for the
associated cyclic unit vector, so that expectations become vector
expectations. The global critical weld of Section~\ref{sec:N} applies in
this representation as well, and the zero-set geometry of
Section~\ref{sec:N} transfers unchanged, since the envelope argument
consumed only the storage properties and the gluing inequality, neither of
which used finite dimension. The full zero set is a strictly increasing
one-to-one graph $P$. The commuting maximizer is supported there;
the transport laws in (S2) allow restriction to a common Borel conull
domain invariant under all finite words in
$a(u)=P^{-1}(-P(u))$ and $\sigma(u)=-u$.

\emph{(S2) Transports without a finite spectrum.} No finite spectral
support is assumed. The transports are expressed through the exact operator blocks
$W=Y(B_3-\tfrac12 I)$ and $W_B=(A_3-\tfrac12 I)V$, which satisfy
$W^2=b(X)^2$ and $WX=-XW$, with the
corresponding $U$-side identities for $W_B$, as projector identities (no
global spectral involution is assumed; it need not exist). Here $X$ and
$U$ are the commuting source and target label operators, and $Y,V$ the
corresponding reflection blocks. These yield exact Radon--Nikodym transport laws for the
scalar spectral measure $\mu$ of the commuting pair $(X,U)$: writing $E$ for
the joint projection-valued measure of that pair, $\mu(\Delta)
=\langle\Omega,E(\Delta)\Omega\rangle$ is a Borel probability measure,
normalized because $\|\Omega\|=1$. Denote its $U$-marginal by $\mu_U$;
graph support transports the $X$-marginal law onto this same measure.

\emph{(S3) Orbits of the response translation.}
The fixed-point set of the response translation $\tau=a\circ\sigma$ carries
no mass --- on it the two transport densities coincide, which fires the
quarter-ceiling elimination against $S>1/4$; the set need not be empty, and
nullity is all that is used. Off it $\tau$ is fixed-point-free and, being a
composition of two decreasing maps, strictly increasing, so each of its
orbits is order-isomorphic to $\mathbb{Z}$ and a Borel transversal can be
\emph{constructed}: select, in a fixed enumeration, the first rational
strictly between the infimum and supremum of the orbit, then its unique
half-open crossing. The rational selection is constant on the entire
orbit. Reflection pairs one or two translation orbits, giving a Borel
transversal for the group in (S4), without an appeal to a general
smoothness assertion.

\emph{(S4) Disintegration, and inherited normalizability.} Pass to
the group $G=\langle a,\sigma\rangle$ generated by the two transports, which
is infinite dihedral and whose orbits may carry nontrivial stabilizers:
$\mu_U$ disintegrates over the countable $G$-orbits. The transport laws are
Radon--Nikodym identities, valid off a single null set of transversal points;
its complement has full measure and is therefore non-empty, and fixing one
point $t$ there, the conditional measure $\mu_t$ satisfies them at every atom
of its orbit. Put $u_n=\tau^n(t)$. Interleaving the weights of this orbit of
$t$ with its reflection $\{-u_n\}$ gives, explicitly,
\[
 c_{2n}=u_n,\qquad c_{2n+1}=-P(-u_n),\qquad
 \widetilde\lambda_{2n}=\sqrt{\mu_t(\{u_n\})},\qquad
 \widetilde\lambda_{2n+1}=\sqrt{\mu_t(\{-u_n\})}.
\]
The odd label and the reflected point supplying its amplitude must not be
identified. Automatically $1\le\sum_j\widetilde\lambda_j^2\le2$:
\emph{normalizability is inherited from the
probability measure, not imposed} --- this is the precise point where the
historical obstruction disappears. Set
$\lambda=\widetilde\lambda/\|\widetilde\lambda\|_2$.
The transport laws \emph{force}
the cocycle $\lambda_{j+1}/\lambda_j=g(c_j)/b(c_j)$, and the Bellman contact
equality closes the Jacobi eigenvalue equation $J\lambda=S\lambda$ exactly,
where $J$ is the Jacobi matrix whose entries (diagonal $\delta$,
off-diagonal $b$) were given in the introduction. Normalization changes
neither the cocycle nor the eigenvalue equation.

\emph{(S5) Installing the eigenvector as a strategy.}
The P\'al--V\'ertesi block identity, in its no-endpoint alternating form,
installs the normalized eigenpair as a spatial strategy with value $S$; the
identity is two-site local, so it applies to the bi-infinite chain
term-by-term, every Bell term being a sum over disjoint two-site blocks with
diagonal sums dominated by $\sum_j\lambda_j^2$ and neighbor sums absolutely
convergent by Cauchy--Schwarz. No limit is taken and no operator topology is
invoked. Every finite principal block of $J$ is itself $\preceq S\,I$, so the
value cannot exceed $S$, and the eigenvalue equation attains it.

Concretely, the bipartite vector is
$\psi_S=\sum_j\lambda_j\,e_j\otimes
e_j$ in the standard basis of
$\ell^2(\mathbb{Z})\otimes\ell^2(\mathbb{Z})$ --- a unit vector by the
rescaling in (S4), as
Theorem~\ref{thm:S} requires; the alternating block form installs
$a_1,a_2,a_3$ on the left factor and $b_1,b_2,b_3$ on the right, each an
orthogonal direct sum of rank-one projections over one of the two perfect
matchings of $\mathbb{Z}$, so that $A_i=a_i\otimes I$ and $B_j=I\otimes b_j$
as Theorem~\ref{thm:S} requires. The block \emph{form}
here is P\'al--V\'ertesi's; the labels $c_j$ and the amplitudes $\lambda_j$
come from the disintegration. This proves Theorem~\ref{thm:S} and, with
Theorem~\ref{thm:N}, the separation in Corollary~\ref{cor:nonclosure}.

\emph{What is machine-checked here.} None of it. The measure-theoretic and
operator-algebraic chains of this section are not formalized in Lean~4, and
the exact-arithmetic guard scripts name the limiting passage, spectral
support and operator closure explicitly among the steps they do not cover.
They are proved here and in the certificate's proof document, and were
audited in the review rounds recorded there (Section~\ref{sec:methods}).

The measure-theoretic steps of this argument --- the conull invariant
set, the Borel transversal, and the uniqueness of the disintegration
(\S\S6--9 of the certificate's proof document) --- were originally
flagged in the certificate's status file as carrying residual proof
risk. An expanded argument was supplied on 7 August 2026 in a full
write-up (25 numbered lemmas, each with complete proof and explicit
quantifier labels, plus an axiom inventory) reviewed in two blind
adversarial rounds: a hostile proof-surface review together with an
independent countermodel search (22 constructed attacks, no
counterexample), followed by a re-review of the repaired document that
verified every repair item by item
(\texttt{AMENDMENT-2026-08-07-SECTIONS-6-9} in the certificate
directory). The status file itself is hash-frozen and deliberately
unamended; the amendment is the correction of record.

\begin{remark}[Envelope versus orbit]
The closure argument of Section~\ref{sec:N} forces the $u\mapsto-u$
symmetry only on the \emph{occupied orbit} of a hypothetical finite
maximizer; nothing forces the critical envelope itself to be
reflection-symmetric, and no step above obtains one wing of the envelope by
reflecting the other. The historical construction's implicit reflection
symmetry is, in retrospect, adjacent to the exact incompatibility that
forced its withdrawal.
\end{remark}

\section{Dimension complexity}\label{sec:rate}

This section determines how fast the finite-dimensional optima approach
$S$. Throughout, $\log$ denotes the natural logarithm; and, in this
section only, $\mu(t)$ denotes the band multiplier (not the spectral
measure of Section~\ref{sec:S}), $\Delta(t)$ the diagonal cost
$\delta(t,t)$ (not a
Borel set). The symbol $I$ keeps its meaning as the identity
operator throughout, and windows of integers are written $\mathcal{I}$.
Define, for each
$d\ge1$,
\begin{align*}
S_d &:= \sup\,\bigl\{\,\langle\psi,\B\,\psi\rangle \;:\;
\dim\mathcal{H}_A\le d,\ \dim\mathcal{H}_B\le d,\ \text{three
projection-valued binary}\\
&\hphantom{{}:={}\sup\,\bigl\{\,}\text{measurements per party on }
\mathcal{H}_A,\ \mathcal{H}_B,\ \psi\in\mathcal{H}_A\otimes\mathcal{H}_B
\text{ a unit vector}\,\bigr\},\\
D(\eps) &:= \min\,\{\, d\in\mathbb{N} \;:\; S-S_d\le\eps \,\}.
\end{align*}
The parameter set is compact --- a finite product of projection varieties
on $\mathbb{C}^d$ together with the unit sphere of
$\mathbb{C}^d\otimes\mathbb{C}^d$ --- and the value is continuous, so the
supremum defining $S_d$ is attained; Theorem~\ref{thm:N} therefore gives
$S_d<S$ for every $d$. $S_d$ is nondecreasing, and $S_d\to S$ by the
construction below, so $D(\eps)$ is well-defined and finite for every
$\eps>0$. (The cap is symmetric on the two parties; the construction
below delivers equal local dimensions, so the same function results.)
The class is projective measurements and pure states, with the dimension
cap imposed on the local Hilbert spaces and no dilation; this loses no generality for the
value: the objective is affine in the state, so an extreme --- pure ---
state attains it, and affine in each binary effect separately, so each
may be replaced by an extreme effect, which is a projection --- the same
replacement as in Section~\ref{sec:N}, and a specifically two-outcome
fact. Naimark dilation is never invoked, so the local dimensions are
unchanged \cite{DGS26}; in particular the POVM-indexed rate is the same
function.

Possible asymptotic scales include $D(\eps)=\Theta(\log(1/\eps))$ ---
geometric convergence; or $D(\eps)/\log(1/\eps)\to\infty$, of which
polynomial dimension cost $D(\eps)\sim\eps^{-c}$ is one instance and the
$2^{\Omega(\eps^{-1/8})}$ of \cite{Coladangelo} a far more expensive one;
or $D(\eps)=o(\log(1/\eps))$, which super-geometric convergence such as
$S-S_d\sim e^{-d^2}$ would produce. The following theorem proves the logarithmic scale, and it takes both halves to do
so: windowing the attaining carrier yields strategies that converge at
least geometrically, and
--- by the lower bound --- nothing converges faster.

\begin{theorem}[Rate]\label{thm:rate}
There are constants $c_1,c_2>0$ with
$c_1\log(1/\eps)\le D(\eps)\le c_2\log(1/\eps)$ for all sufficiently
small $\eps>0$. The upper half holds with the explicit derived constant:
$D(\eps)\le 23.9010650\,\log(1/\eps)$ for all sufficiently small
$\eps$.
\end{theorem}

\noindent\emph{Where the two halves are proved.} The lower proof is
in \S\ref{sec:rate-lower}, and the constructive upper proof is in
\S\ref{sec:rate-upper}. Both use the fixed critical storage and compact
interiority derived below. Only the upper proof consumes the spatial
carrier and its strict graph structure.

In gap form, given the monotonicity of $S_d$:
there is $\keff>0$ (defined in \S\ref{sec:rate-upper}, with
$1/\keff\le23.9010650$) such that for every $0<\alpha<\keff$ there is $d_0$
with $S-S_d\le e^{-\alpha d}$ for all $d\ge d_0$; and there are $c,C>0$
with $S-S_d\ge c e^{-Cd}$ for every $d$.

\subsection{The lower bound by finite weighted flow}\label{sec:rate-lower}

We prove, with constants depending only on the fixed Bell functional,
\begin{equation}\label{eq:lb}
 S-S_d\ge c_* e^{-C_*d}\qquad(d\ge1).
\end{equation}
The proof uses neither a dimension-preserving conversion to a chain nor
identification of vectors in different joint spectral cells. There can be
quadratically many such cells. We count only the distinct local spectral
labels, and control all nonnegative cell weights simultaneously.

\subsubsection*{Critical data and the original-dimension interface}

Write
\[
 \delta(x,u)=xu+(x-u)/2-1,\quad b(t)=\tfrac12\sqrt{1-t^2},
 \quad w=b^2.
\]
The critical-storage construction of Section~\ref{sec:setup} gives a
concave $3/2$-Lipschitz function $g$ on $[-1,1]$, positive in the open
interval, such that
\begin{equation}\label{eq:wf-feasibility}
 \frac{w(x)}{g(x)}+g(u)\le S-\delta(x,u),\qquad
 g(t)g(-t)\ge w(t).
\end{equation}
Initially the quotient is taken at interior points. To make the inherited
input precise, for $q>S$ take the infimum $g_q$ of terminal Schur pivots
over finite Jacobi histories. The bound $J\preceq SI$ gives pivots at least
$q-S$; extending a history gives the first inequality at level $q$.
Each terminal-pivot function has slope $1/2-x$ in the terminal label, so
their infimum is concave and $3/2$-Lipschitz. Monotonicity in $q$ and the
common modulus give a uniform limit $g$ as $q\downarrow S$.
Feasibility with a fixed target label bounds $w(t)/g_q(t)$ uniformly for
each interior $t$, proving $g(t)>0$ there. Concatenating a history ending
at $t$ with the reflected reverse of a history ending at $-t$ gives two
central Schur pivots coupled by $b(t)$. The resulting positive
$2\times2$ matrix gives the second inequality in
\eqref{eq:wf-feasibility} after taking the two infima and the limit.
This is the critical-storage input, not a consequence of the old
finite-dimensional rate argument.
For a related treatment of canonical storage and positive welds for
reversible Bell functionals, see Mghirbi \cite{MghirbiDynamics}.
The argument below supplies the unrestricted lower bound directly;
no such bound is imported from that work.

Here is an elementary endpoint check using only $S<1/3$.
One-edge histories give $g(1),g(-1)\le S$. If $g(1)=0$, Lipschitz
continuity and reflection gluing imply, for $-1<t<1$,
\[
 g(t)\le\tfrac32(1-t),\qquad
 g(-t)\ge\frac{1-t^2}{4g(t)}\ge\frac{1+t}{6}.
\]
Letting $t\uparrow1$ contradicts $g(-1)\le S<1/3$. Reflecting proves
$g(-1)>0$ also. Thus $m:=\min_{[-1,1]}g>0$, and all quotients below
extend continuously to the endpoints.

Set
\[
 p=w/g,\quad \alpha(t)=b(t)\sqrt{g(-t)/g(t)},\quad
 \beta(t)=\alpha(-t),\quad
 \phi(x,u)=S-\delta(x,u)-\alpha(x)-\beta(u).
\]
Feasibility and its reflected version, followed by Cauchy--Schwarz, give
$\phi\ge0$. In detail, the two vectors
$(\sqrt{p(x)},\sqrt{g(u)})$ and
$(\sqrt{g(-x)},\sqrt{p(-u)})$ have inner product
$\alpha(x)+\beta(u)$ and squared norms at most $S-\delta(x,u)$.
At $x=\pm1$ they cannot be proportional: the first coordinate of the
first vector vanishes, whereas both $g(u)$ and $g(-x)$ are positive.
Consequently $\phi>0$ on those two boundary edges; reflection treats the
other two. Compactness supplies a symmetric interval
$K=[-a,a]\subset(-1,1)$ containing every zero strictly inside $K^2$,
and a constant $a_0>0$ with
\begin{equation}\label{eq:wf-collar}
 \inf_{(x,u)\notin K^2}\phi(x,u)\ge a_0.
\end{equation}
For example, minimize on the closed boundary collar
$\max(|x|,|u|)\ge a$. No minimum on a nonclosed set is needed.

First reduce arbitrary binary effects and mixed states without dimension
loss. At fixed state the Bell value is affine in each effect; successively
maximize each coordinate at a projection, then choose a top eigenvector
of the resulting Bell operator. The value cannot decrease and neither
local dimension changes. It therefore suffices to prove the lower bound
for projective pure-state strategies. For such measurements on the
original tensor product, put
\[
 X=A_1+A_2-I,\quad Y=A_2-A_1,\quad
 U=B_1+B_2-I,\quad V=B_2-B_1,
\]
and $W_A=Y(B_3-I/2)$, $W_B=(A_3-I/2)V$. Direct multiplication, using
idempotence of the measurement projections, gives
\[
 W_A^*=W_A,\quad W_AX=-XW_A,\quad W_A^2=b(X)^2,
\]
and the analogous identities for $W_B,U$. In particular $W_A$ is not
assumed to commute with $U$. The operators
$Q_A=\alpha(X)-W_A$ and $T_A=\alpha(-X)+W_A$ are self-adjoint and
satisfy $Q_AT_A=0$: use $W_Af(X)=f(-X)W_A$ and
$\alpha(t)\alpha(-t)=b(t)^2$. They therefore commute. Their sum is
nonnegative, so their joint spectral values show that both factors are
nonnegative. Likewise $Q_B=\beta(U)-W_B\ge0$. Expanding gives
\begin{equation}\label{eq:wf-weld}
 SI-\B=\phi(X,U)+Q_A+Q_B.
\end{equation}
For a unit vector of deficit $\eps=S-\langle\B\rangle$ and its joint
spectral probability measure $\mu$, positivity and bounded norms imply
\begin{equation}\label{eq:wf-residual}
 \int\phi\,d\mu\le\eps,\qquad
 \|Q_A\psi\|+\|Q_B\psi\|\le C\sqrt\eps,\qquad
 \mu((K^2)^c)\le a_0^{-1}\eps.
\end{equation}
Indeed $\|Q_A\psi\|^2\le\|Q_A\|\langle Q_A\rangle$, with a fixed,
dimension-independent norm bound. This derives the approximate response
estimate from the positive weld, not from an exact-equality statement.

If the original local dimensions are at most $d$, each of $X,U$ has at most $d$ distinct
eigenvalues, regardless of spectral multiplicities.

\subsubsection*{A quantitative monotone predictor}

Put $C_0(x)=S+1-x/2-p(x)$ and let $H$ be its greatest convex minorant
on the full interval $[-1,1]$. There is a fixed $M\ge1$ such that
$h=H'$ is nondecreasing and $M$-Lipschitz in the interior, with a
continuous constant extension past either endpoint.
For completeness, for smooth positive concave approximants to $g$,
\[
 p''=\frac{w''}{g}-\frac{2w'g'}{g^2}
       +\frac{2w(g')^2}{g^3}-\frac{wg''}{g^2}
       \ge-\frac1m-\frac{4L}{m^2},
\]
where $L$ bounds the slopes and the approximants are at least $m/2$.
Concave affine extensions followed by mollification and uniform limits
show that $C_0$ is semiconcave. One may take
$M=1+1/m+4L/m^2$. On every open gap where $H<C_0$, maximality makes
$H$ affine. At an interior contact,
\[
 C'_{0,-}\le H'_-\le H'_+\le C'_{0,+},
\]
whereas semiconcavity reverses the outside inequality. All four values
coincide. Adding the semiconcavity inequalities at contact points $x<y$
gives $0\le H'(y)-H'(x)\le M(y-x)$. The affine gaps extend this estimate
to arbitrary points, establishing the assertion.

The affine minorant at $\min_z(C_0(z)-uz)$ lies below $H$, so
\[
 g(u)-u/2\le\min_z(C_0(z)-uz)=\min_z(H(z)-uz).
\]
Write $D=S-\delta(x,u)$, $L_1=D-p(x)-g(u)$ and
$L_2=D-g(-x)-p(-u)$. Both are nonnegative, and
\[
 \alpha(x)+\beta(u)\le\sqrt{(D-L_1)(D-L_2)}
 \le D-(L_1+L_2)/2.
\]
Consequently
\begin{equation}\label{eq:wf-envelope-gap}
 H(x)-ux-\min_z(H(z)-uz)\le L_1\le2\phi(x,u).
\end{equation}
Choose $\rho>0$ with every $x\in K$ at least $\rho$ from the interval
endpoints. For $F(z)=H(z)-uz$, the admissible step
$z=x-\rho\operatorname{sign}F'(x)$ shows that a gap smaller than
$M\rho^2/2$ implies $|F'(x)|<M\rho$. The full step $x-F'(x)/M$ is
then admissible; the smoothness inequality gives a decrease at least
$|F'(x)|^2/(2M)$. Thus, for every $s$ below a fixed sufficiently small
positive threshold,
\begin{equation}\label{eq:wf-predictor}
 x,u\in K,\ \phi(x,u)\le s
 \quad\Longrightarrow\quad |u-h(x)|\le2\sqrt{Ms}.
\end{equation}

\subsubsection*{Scalar balance without identification of joint cells}

Let $p_X,p_U$ be the original spectral marginals, and put
$r(t)=\sqrt{g(-t)/g(t)}$, $R(t)=r(t)^2$. On $K$, $R$ is Lipschitz,
with $0<R_{\min}\le R\le R_{\max}<\infty$.
The operator $b(X)^{-1}W_A$ is an isometry on the $X$-corridor and
reflects its spectral projections. Projecting the response residual in
\eqref{eq:wf-residual}, dividing by $b(t)\ge b_0>0$, and using the
reverse triangle inequality yields
\[
 \sum_{t\in K}
 \left(\sqrt{p_X(-t)}-r(t)\sqrt{p_X(t)}\right)^2\le C\eps.
\]
Missing spectral atoms have weight zero. Cauchy--Schwarz converts this
to the total $\ell^1$ bounds
\[
 \|\mathcal Fp_X-Rp_X\|_{1,K}\le C_1\sqrt\eps,\qquad
 \|\mathcal Fp_U-R^{-1}p_U\|_{1,K}\le C_2\sqrt\eps,
\]
where $\mathcal F$ reflects a measure. These inequalities use genuine
projected response vectors but not their localization in joint cells.

Average measures only:
$\bar\mu=(\mu+J_*\mu)/2$, $J(x,u)=(-u,-x)$. Since
$\phi\circ J=\phi$, the defect bound is unchanged. Its marginals
$p=(p_X+\mathcal Fp_U)/2$, $q=(p_U+\mathcal Fp_X)/2$ satisfy
\[
 q-Rp=\tfrac12(\mathcal Fp_X-Rp_X)
       -\tfrac12R(\mathcal Fp_U-R^{-1}p_U).
\]
Fix the vertex universe to be the union of the original and reflected
spectral labels in $K$, of cardinality $n\le4d$. Retain only cells
$(x,u)\in K^2$ with $\phi(x,u)\le s\le1$, as directed edges $x\to u$.
Keep isolated vertices. If $m_s$ is retained mass and $o,i$ are outgoing
and incoming masses, Markov's inequality gives fixed constants with
\begin{equation}\label{eq:wf-balance}
 1-m_s\le D_0\eps/s,\qquad
 \|i-Ro\|_1\le C_b\sqrt\eps+(1+R_{\max})D_0\eps/s.
\end{equation}
No physical realization of $\bar\mu$ is asserted or needed.

\subsubsection*{Cycles and paths}

An elementary monotonicity observation will be used. If
$x_0,\ldots,x_k=x_0$ satisfies $|x_{j+1}-h(x_j)|\le\delta$ with
$h$ nondecreasing, then $|h(t)-t|\le\delta$ throughout the visited
interval. Every interior level is crossed upwards and downwards; an
upward crossing gives $h(t)\ge t-\delta$, and a downward one gives
$h(t)\le t+\delta$. Predecessor and successor inequalities give the
same bounds at the minimum and maximum. In particular every cycle edge
has length at most $2\delta$.

Put $\Delta=S-1/4>0$. Since $t^2-1+2b(t)\le1/4$, on the diagonal
\[
 \phi(t,t)=S-[t^2-1+b(t)(r(t)+r(t)^{-1})].
\]
For $|R(t)-1|\le c\le1/2$ the identity
$r+r^{-1}-2=(R-1)^2/[r(r+1)^2]$ implies
$\phi(t,t)\ge\Delta-c^2$. Choose
$c=\min(1/2,\sqrt\Delta/2)$. On a retained cycle,
\eqref{eq:wf-predictor} instead bounds
$\phi(x,x)\le s+2L_\phi\delta$, where
$\delta=2\sqrt{Ms}$ and $L_\phi$ is a Lipschitz constant on $K^2$.
Choose a fixed sufficiently small $\delta_*>0$ so that the latter
quantity is at most $\Delta/2$ whenever $\delta\le\delta_*$.
Every cyclic vertex must then have $R\le1-c$ or $R\ge1+c$.
Every internal edge of a cyclic strongly connected component belongs
to a cycle. Decreasing $\delta_*$ so that $2L_R\delta_*<2c$ shows
that such a component is entirely low or entirely high.

Individual components are not enough: a low cycle feeding a high cycle
can support a nonzero exactly balanced flow. To exclude this, also take
$\delta_*\le c/(4\max(1,L_R))$, and choose the cut using the previously
fixed universe:
\begin{equation}\label{eq:wf-cut}
 \delta_n=\delta_*(M+1)^{-n},\qquad s_n=\delta_n^2/(4M).
\end{equation}
Starting at a cyclic vertex $a$, any $k$-edge path satisfies
\[
 |x_{j+1}-a|\le M|x_j-a|+2\delta_n,\qquad
 |x_k-a|\le2\delta_n\sum_{j=0}^{k-1}M^j
              \le2\delta_n(M+1)^k.
\]
A shortest path is simple, hence has at most $n-1$ edges. Its endpoint
cannot be an opposite-gain cyclic vertex: such vertices differ in $R$
by at least $2c$, while the displayed distance and Lipschitz continuity
give a difference at most $c/2$. In particular no low cyclic component
can reach a high cyclic component.

\subsubsection*{A signed potential that controls every branch}

For any $n$-vertex directed graph with these properties there is a real
function $f$ such that
\begin{equation}\label{eq:wf-potential}
 R_xf_x-f_y\ge1\quad(x\to y),\qquad
 \|f\|_\infty\le B_*^{n+1},\quad
 B_*=4+2R_{\max}+2/R_{\min}+2/c.
\end{equation}
Here is the construction. Let $\mathcal U$ be all vertices reachable
from low cyclic components. It is forward closed and contains no high
cycle. Its complement is backward closed, and no strongly connected
component straddles the split: any component meeting $\mathcal U$ is
contained in it by forward closure. On its condensation directed acyclic graph choose constants in
topological order,
\[
 g_C=\max\{1,1/c,\max_{D\to C}(R_{\max}g_D+1)\},
\]
omitting empty inner maxima, and put $f=-g_C$ on $C$. An internal
edge in a cyclic component has $(1-R_x)g_C\ge1$; edges between
components satisfy $g_{C(y)}-R_xg_{C(x)}\ge1$ by construction.
An acyclic singleton has no internal edge. On the complementary
condensation graph process in reverse topological order and set
\[
 f_C=\max\{1/R_{\min},1/c,
       \max_{C\to D,\ D\not\subset\mathcal U}(f_D+1)/R_{\min}\}.
\]
Its cyclic components are high, so internal edges obey
$(R_x-1)f_C\ge1$; edges between components obey the same required
inequality by the recursion. Edges into $\mathcal U$ obey it because
$f_x\ge1/R_{\min}$ and $f_y<0$. No edge leaves $\mathcal U$.
At most $n$ stages, with fixed bounded growth at each, give the norm
bound in \eqref{eq:wf-potential}.

For arbitrary nonnegative weights on all retained edges, summing gives
\begin{equation}\label{eq:wf-all-branches}
 m_s\le\sum_{x\to y}\bar\mu_{xy}(R_xf_x-f_y)
       =\sum_x f_x(R_xo_x-i_x)
       \le B_*^{n+1}\|Ro-i\|_1.
\end{equation}
This is why branching and spectral multiplicity introduce no missing
payment: no edge, row, or component is replaced by a selected path.

\subsubsection*{Assembly}

If the fixed universe is empty, the collar estimate already gives a
positive lower bound on $\eps$. Otherwise combine
\eqref{eq:wf-balance}--\eqref{eq:wf-all-branches}, with
$C_3=(1+R_{\max})D_0$ and constants enlarged to at least one:
\[
 1-D_0\eps/s_n\le B_*^{n+1}
                 (C_b\sqrt\eps+C_3\eps/s_n).
\]
Either $D_0\eps/s_n\ge1/2$, or one of the two right-hand terms is
at least $1/4$. Hence
\[
 \eps\ge\min\left\{\frac{s_n}{2D_0},
 \frac{1}{16C_b^2B_*^{2n+2}},
 \frac{s_n}{4C_3B_*^{n+1}}\right\}.
\]
By \eqref{eq:wf-cut} this is at least $c_0e^{-C_0n}$ with fixed positive
constants. Since $n\le4d$, \eqref{eq:lb} follows. It also excludes exact
finite-dimensional attainment without using spatial attainment or the
old common-return argument. Finally,
$\eps\ge c_*e^{-C_*d}$ implies
$d\ge C_*^{-1}\log(c_*/\eps)$, proving the lower half of
Theorem~\ref{thm:rate}.\hfill$\square$

\paragraph{Verification scope.}
This proof is analytic. Exact finite-graph tests exercise the signed
potential and negative controls, but do not prove the analytic estimates.
The Lean module \texttt{I3322Kernel.WeightedFlow} verifies the finite
all-edge summation identity, the residual bound conditional on an edge
potential, three local potential inequalities, and the elementary
assembly alternative. It does not prove the existence of the potential
from strongly connected components, convex-envelope regularity, or the
quantum-to-flow reduction. The full chain remains an analytic proof.
The earlier staircase and selected-walk cores are not dependencies of
this replacement.

\subsection{The upper bound, explicitly}\label{sec:rate-upper}

The upper half is constructive, and is given here in compressed form: the
construction, the constants, and the route are all stated, while the full
accounting of the endpoint-diagonal payment and its constants is carried
by the proof document \cite{RUB26}. It consumes
Theorem~\ref{thm:S} and the inputs named below; the reductions, the exact
statements of the transported inputs, and the assembly are not formalized
(\S\ref{sec:rate-scope}).

\emph{The carrier.} The proof of Theorem~\ref{thm:S} supplies a positive
unit vector $\lambda\in\ell^2(\mathbb{Z})$ --- the \emph{carrier} --- and
a bi-infinite label sequence $(c_j)\subset(-1,1)$ with:
\begin{itemize}
\item[(i)] the certificate's interior zero locus $Z$ (the zero set of its
scalar weld functional) is the graph of a strictly increasing one-to-one
Borel map $P$, and $P(c_{j+1})=c_j$ for every $j$ --- one map, every link
(Theorem~\ref{thm:S} certificate, \S\S6 and~10);
\item[(ii)] the transport law $\lambda_{j+1}/\lambda_j = g(c_j)/b(c_j)$
(Theorem~\ref{thm:S} certificate, \S11), where $b(t)=\sqrt{1-t^2}/2$ and $g$ is the critical
storage of Section~\ref{sec:N}, continuous and positive on the open
interval (a uniform positive bound on the compact label set follows
below);
\item[(iii)] the eigen-equation $J\lambda=S\lambda$ for the Jacobi matrix
$J_{jj}=\delta(c_{j-1},c_j)$, $J_{j-1,j}=b(c_{j-1})$, where
$\delta(x,u):=xu+(x-u)/2-1$ is the certificate's diagonal cost;
\item[(iv)] every adjacent label pair lies in $Z$, and $Z$ is compactly
interior, $Z\subset\subset(-1,1)^2$.
\end{itemize}
Item~(i) is proved in the certificate's \S\S6--9 (conull invariant set,
Borel transversal, uniqueness of disintegration) and \S10; an expanded
write-up of \S\S6--9 is included in the same certificate directory as
\texttt{AMENDMENT-2026-08-07-SECTIONS-6-9}. The compact interiority~(iv)
is the endpoint-positivity statement of \cite{Repo}, used here exactly as
established there and with the status recorded in
\S\ref{sec:rate-scope}. The positivity in~(ii), by contrast, needs no
such input on the closed interval: what the argument below uses is a
positive lower bound on the compact label set only, and that follows
from Section~\ref{sec:N}'s own properties of $g$ --- continuous,
positive on the open interval --- once (iv) confines the labels to a
compact subset of it. Let $K$ be the union of the two coordinate
projections of the closure of $Z$; by (iv) it is a compact subset of
$(-1,1)$;
all labels and label limits lie in $K$, and on $K$ the functions $b$,
$g$, $r:=g/b$ are continuous with $b\ge b_0>0$, $g\ge m_g>0$, and $r$,
$1/r$ bounded by $R_{\max}<\infty$.

\emph{Monotone labels and geometric tails.} Since $P$ is strictly
increasing and one-to-one with $P(c_{j+1})=c_j$ for every $j$, the labels
satisfy $c_{j+1}=P^{-1}(c_j)$ and $c_{j-1}=P(c_j)$: the orbit is
functional in both directions, so it is strictly increasing, strictly
decreasing, or constant, and constancy at a single adjacent pair
propagates to the whole sequence. The constant case is excluded: a
constant label $c$ would make $\lambda$ exactly geometric with ratio
$r(c)$ --- not in $\ell^2(\mathbb{Z})$ in one direction if $r(c)\neq1$,
and a non-normalizable constant vector if $r(c)=1$. Hence the whole
sequence is strictly monotone, confined to the compact $K$, so both tail
limits $t_\pm\in K$ exist, and by the transport law and continuity of $r$
the outward one-step ratios converge:
\[
y_+ := \lim_{j\to+\infty}\lambda_{j+1}/\lambda_j = r(t_+),\qquad
y_- := \lim_{j\to-\infty}\lambda_{j-1}/\lambda_j = 1/r(t_-).
\]
Dividing row $j$ of $J\lambda=S\lambda$ by $\lambda_j$ and passing to
either limit gives
\begin{equation}\label{eq:band}
y+1/y \;=\; \mu(t) := \frac{S-\Delta(t)}{b(t)},\qquad
\Delta(t):=\delta(t,t)=t^2-1,
\end{equation}
at $t=t_\pm$. The elementary band identity is, for $s=\sqrt{1-t^2}$,
\begin{equation}\label{eq:bandid}
\Delta(t)+2b(t) \;=\; t^2-1+s \;=\; s(1-s) \;\le\; 1/4
\end{equation}
(machine-checked: \texttt{band\_identity},
\texttt{s\_mul\_one\_sub\_s\_le\_quarter},
\texttt{band\_quarter\_ceiling}). Together with $b\le1/2$
(\texttt{amplitude\_b\_le\_half}) and the certified lower window endpoint
$S_{\mathrm{LO}}:=0.2508753845015185<S$ it yields, for every interior $t$,
\begin{equation}\label{eq:mumin}
\begin{aligned}
\mu(t)-2 \;&=\; (S-\Delta-2b)/b \;\ge\; (S-1/4)/b \;\ge\; 2(S-1/4)\\
&>\; 2(S_{\mathrm{LO}}-1/4) \;=\; 0.001750769003037,\\
\text{so }\ \mu(t) \;&\ge\; \mu_{\min} := 2.001750769003037.
\end{aligned}
\end{equation}
Since $\ell^2$ excludes the growing root of \eqref{eq:band} at each end,
both outward ratios equal the decaying root
$x_{\mathrm{dec}}(\mu)=2/(\mu+\sqrt{\mu^2-4})$, strictly decreasing in
$\mu$, so uniformly
\begin{equation}\label{eq:ydec}
y_\pm \;\le\; x_{\mathrm{dec}}(\mu_{\min}) \;\le\; 0.9590241,\qquad
y_\pm^2 \;\le\; 0.91972725 \;<\; 1.
\end{equation}
In particular $r(t_+)=y_+<1$ while $r(t_-)=1/y_->1$: the transport ratio
crosses $1$ strictly between the two tail limits --- a nontrivial
corroboration of the two-ended open-chain picture.
Tail mass sums amplitude squares, so with $\kappa_\pm:=-2\log y_\pm$ the
tails obey $\lambda_j^2=\exp(-\kappa_\pm|j|+o(|j|))$ at the
respective ends. In particular, for every $0<\eta<\min(\kappa_-,\kappa_+)$
there is $A_\eta<\infty$ with
$\lambda_j^2\le A_\eta e^{-(\kappa_\pm-\eta)|j|}$ on the corresponding
tail. Ratio convergence alone does not justify a fixed prefactor at the
limiting exponent. Carrying the down-rounded surrogate
$K_0 := 0.08367827985 \le -2\log x_{\mathrm{dec}}(\mu_{\min})$ (computed
against the exact $x_{\mathrm{dec}}(\mu_{\min})$, not the rounded display
in \eqref{eq:ydec}, which is too weak to support it), which is safe both
in $K_0/2$ (rounded down) and in the denominator of
$2/K_0$ (a down-rounded denominator over-estimates the quotient, the safe
direction for an upper display):
\begin{equation}\label{eq:kappa}
\begin{aligned}
\kappa_\pm \;&\ge\; K_0 = 0.08367827985 \ \text{per index},\\
\keff \;&:=\; (1/\kappa_- + 1/\kappa_+)^{-1} \;\ge\; K_0/2 \;\ge\;
0.0418391,\\
1/\keff \;&\le\; 2/K_0 = 23.90106493088\ldots \;\le\; 23.9010650.
\end{aligned}
\end{equation}
(All displayed decimals are bounds rounded in the safe direction and
re-verified in exact rational arithmetic by the repository's guard
scripts, which re-derive every displayed constant from its stated inputs
and fail on any mismatch. Because $1/\keff\le 2/K_0 =
23.90106493088\ldots$ sits strictly below $23.9010650$ with a genuine
gap, the eventual bound $D(\eps)\le 23.9010650\cdot\log(1/\eps)$ follows
--- a bare $\limsup\le 23.9010650$ alone would not give it.)

\emph{The truncation, at exact local dimension $d$.} For a window
$\mathcal{I}=[-L,R]\cap\mathbb{Z}$, keep every full $2\times2$ rank-one block of
each of the carrier's two alternating matchings whose paired indices lie
in $\mathcal{I}$; at a severed endpoint pair, replace the block by the
one-dimensional projector on the retained index (either of the two
admissible one-dimensional assignments serves; we fix this choice, and the
choice is part of the endpoint-diagonal payment accounted below). All
six measurement
operators remain exact projections --- each is a direct sum of orthogonal
rank-one blocks and one-dimensional projectors (verified symbolically at
arbitrary block angles for window sizes $m=3,\ldots,8$ by an
independently written script; the general case is the direct-sum
structure itself) --- and the local dimension is exactly $d=|\mathcal{I}|$, with no
dilation and no padding. The truncated state is the renormalized window
of $\lambda$.

\emph{Error accounting.} Let $v_{\mathcal{I}}$ be the value of the principal Jacobi
compression on $\mathcal{I}$ and $V_{\mathcal{I}}$ the Bell value of the truncated strategy
just built. The exact two-boundary flux identity (two lines from
$J\lambda=S\lambda$):
\[
S-v_{\mathcal{I}} \;=\; B_{\mathcal{I}}/M_{\mathcal{I}},\qquad
B_{\mathcal{I}} := b(c_{-L-1})\lambda_{-L-1}\lambda_{-L}
+ b(c_R)\lambda_R\lambda_{R+1},
\]
with $M_{\mathcal{I}}=\sum_{j\in\mathcal{I}}\lambda_j^2$ the window mass --- only the two cut
edges are unpaid. Relative to the compression, the completion changes
exactly two kinds of term: the two cut-bond off-diagonals (already and
only accounted for by the flux identity --- not charged twice) and the two
retained endpoint diagonals; nothing else changes. Each of the eight joint
Bell terms has coefficient of modulus $1$, and a joint term's diagonal is
a product of two projector diagonals, so its change has modulus at most
$2$; the marginal coefficients contribute their absolute sum $4$; hence
the safe count $C_{\mathrm{diag}}\le 4\cdot1+8\cdot2=20$ and
\[
0 \;\le\; S-V_{\mathcal{I}} \;\le\; (1+20R_{\max}/b_0)\,
B_{\mathcal{I}}/(1-T_{\mathcal{I}}),
\]
with $T_{\mathcal{I}}:=1-M_{\mathcal{I}}$ the omitted tail mass.
Indeed, the ratio bounds and $b\ge b_0$ give
\[
\lambda_{-L}^2+\lambda_R^2\le (R_{\max}/b_0)B_{\mathcal{I}}.
\]
The flux contributes $B_{\mathcal{I}}/M_{\mathcal{I}}$, while the two
endpoint diagonals contribute at most
$20(\lambda_{-L}^2+\lambda_R^2)/M_{\mathcal{I}}$.
This also distinguishes the coupling-weighted flux used here from the
unweighted boundary product used in \cite{RUB26}; their prefactors must
not be interchanged. The fixed prefactor does not change the asymptotic
exponent. Both $B_{\mathcal{I}}$ and $T_{\mathcal{I}}$ are bounded by
$e^{-\kappa_-L+o(L)}+e^{-\kappa_+R+o(R)}$, so
balancing $L:R$ as $\kappa_+:\kappa_-$ with $|\mathcal{I}|=d$ exactly:
\begin{equation}\label{eq:ub}
S-S_d \;\le\; S-V_{\mathcal{I}_d} \;\le\; \exp(-\keff\cdot d + o(d)),
\end{equation}
where the $o(d)$ absorbs the multiplicative constants, the integer
rounding of $L$ and $R$, and the ratio-to-tail conversion; therefore, for
every $0<\eta<\keff$,
$D(\eps)\le(\keff-\eta)^{-1}\log(1/\eps)$ eventually; letting
$\eta\downarrow0$ gives
$\limsup D(\eps)/\log(1/\eps)\le 1/\keff\le 23.9010650$. This proves
the upper half of Theorem~\ref{thm:rate}; with the new lower proof in
\S\ref{sec:rate-lower},
Theorem~\ref{thm:rate} follows.\hfill$\square$

\begin{remark}[Concrete dimensions]
The geometric regime is visible at concrete sizes: the repository's
independently written verification script evaluates explicit strategies
of the P\'al--V\'ertesi family --- a lower bound on $S_d$, not the
truncation above --- on the sampled ladder
$d\in\{3,4,6,8,12,16,24,33\}$, with values exceeding the two-qubit
ceiling $1/4$ at $d=24$ and $d=33$ (high-precision evaluation with
rigorous Rayleigh-quotient lower bounds reported by the script); by
monotonicity of $S_d$, $S_d>1/4$ for all $d\ge24$. The script's fixed
parametrization is not P\'al--V\'ertesi's tuned family, so no conclusion
about $d<24$ follows from these evaluations, and nothing is claimed about
the exact crossing point; in particular there is no tension with
P\'al--V\'ertesi's own dimension-scan report, which concerns their tuned
family.
\end{remark}

\subsection{Scope, and what is not claimed}\label{sec:rate-scope}

\emph{Constants.} The constants in the $\Theta$ are existential. The
single explicit number $23.9010650$ bounds the upper half's constant
only; it derives from the certified window and band algebra via
an exact-rational chain, and is not claimed to be sharp.
The lower half's constants $c,C$ are
existential and no explicit values are claimed, so no explicit two-sided
window on $D(\eps)/\log(1/\eps)$ is asserted; the $o(d)$ in
\eqref{eq:ub} is not made explicit, and the Theorem's inequalities are
claimed for all sufficiently small $\eps$.

\emph{Machine-checking boundary.} Existing Lean theorems cover selected
scalar and combinatorial facts in the earlier development. They do not
formalize the new predictor, quantum response reduction, or existence of
the signed potential. The elementary finite assembly alternative is
formalized, but the complete analytic assembly is not.
Exact graph checks are controls, not proofs of these
analytic implications.

\emph{Inheritance.} Both rate bounds use the critical storage and its
compactly interior zero locus. Section~\ref{sec:rate-lower} derives the
endpoint positivity, collar, and approximate response estimates directly
from that storage, reflection gluing, and $1/4<S<1/3$. The lower proof
requires no strict graph structure and no spatial carrier. The upper
proof still consumes the spatial-attainment certificate and its amended
graph and orbit construction. This new lower proof neither revalidates
those separate dependencies nor consumes the withdrawn paired-block
step in the earlier rate certificate \cite{RLB26}.

\emph{Priority, stated against the nearest prior art.} Two
qualifications are owed. First, the $\Omega$-type dimension bounds already in the
literature --- for instance the $2^{\Omega(\eps^{-1/8})}$ of
\cite{Coladangelo} --- attach to witnesses of the \emph{distinct}
separation $\Cqs\neq C_{qa}$, and \cite{Coladangelo} gives a lower bound
only, with no matching upper bound or tightness claim. Second, and more
directly: Coladangelo and Stark remark, in the arXiv version of
\cite{ColadangeloStark} immediately following their non-attainability
theorem, that repeated application of their Schmidt-orbit maps exhibits
an infinite geometric sequence and that ``this can be used to obtain some
quantitative bounds on the dimension required to induce a correlation
close to the ideal one''; they decline to prove it, on the stated ground
that more useful bounds exist for the $\Cqs\neq C_{qa}$ separation. That
remark does not appear in the published version. We take it at face
value: for their ideal state, whose Schmidt coefficients are exactly
geometric, an $O(\log(1/\eps))$ upper bound is a routine truncation
argument, and their own discussion already names truncation to
bounded-energy subspaces as the approximation mechanism. What we claim is
therefore specific: a two-sided characterization for $I_{3322}$,
with the upper half in \S\ref{sec:rate-upper} and the unrestricted
lower half in \S\ref{sec:rate-lower}. We note also what their
method does not immediately give: the present carrier is \emph{not}
geometric but asymptotically geometric at two different rates at its two
ends, which is why the constant is the harmonic-type combination
$\keff=(1/\kappa_-+1/\kappa_+)^{-1}$ rather than a single decay rate ---
that two-ended structure, forced by $r(t_+)<1<r(t_-)$, is the technical
content of \S\ref{sec:rate-upper}. These comparisons distinguish the scopes of the results; they are not
a proof of an exhaustive priority claim.

\section{Discussion: attainment and open carriers}\label{sec:discussion}

The two theorems exhibit a sharp mechanism: \emph{finiteness forces the
equality transports to close (return), and closure caps the value strictly
below $S$; the value above the closed ceiling is carried only by an infinite
open chain}. It is natural to ask how general this picture is. In every
case known to us, a Bell functional with proven finite-dimensional
attainment of its global supremum closes for an identifiable algebraic
reason: bipartite correlator (XOR) functionals --- including the chained
inequalities --- via Tsirelson's Clifford construction \cite{Tsirelson,CHTW}
(anticommutation is two-step closure); all $(2,2;2,2)$ functionals, whose
extreme points are realized by qubit strategies \cite{Masanes}; fixed CGLMP
functionals, whose global optimum is attained at local dimension $d$ (proven
for $d=3,4$ \cite{IoannouRosset}; we make no minimality claim); and,
informally, those families whose sum-of-squares certificates force
finite-dimensional algebraic relations on optimal strategies.

Conversely,
both known explicit transport-class nonattainment objects are infinite open
ladders: the Coladangelo--Stark state is an exactly geometric ladder whose
nonattainment proof (Theorem~12 of the arXiv version of
\cite{ColadangeloStark}) is an explicit
no-return orbit argument on the Schmidt spectrum,
and the P\'al--V\'ertesi family is a half-infinite Jacobi chain with offset
$2\times2$ blockings \cite{PalVertesi} --- the same architecture, on smaller
alphabets. Coladangelo and Stark suggested that the block-diagonal form of
the conjectured optimal $I_{3322}$ measurements, resembling that of their
own construction, makes the inequality ``potentially amenable to ideas and
techniques from our work''; Theorem~\ref{thm:N} may be read as an answer to
that suggestion, reached by a transfer-operator route. Embezzlement-based
nonattainment \cite{vanDamHayden,JLV,Coladangelo} is carried by the
log-flat ladder $p_j\propto1/j$ --- an open chain of a different decay
class; whether this heuristic correspondence with the operator-algebraic
classification of embezzlement \cite{vLSWW} can be made precise is an open
question we do
not pursue here.

Group-theoretic and type-based obstructions
\cite{Slofstra1,DPP,MusatRordam,MIPRE} form a third family outside this
transport picture: Slofstra's undecidability of perfect-strategy existence
for linear-system games \cite{Slofstra1}, the RE-completeness of
approximating the quantum value \cite{MIPRE}, and the arithmetical-hierarchy
separation of \cite{MNY}, in which deciding whether the quantum value of a
two-player nonlocal game equals $1$ is $\Pi_2$-complete
(Mousavi--Nezhadi--Yuen) whereas deciding whether the commuting-operator
value equals $1$ is $\Pi_1$ (Slofstra, as restated there), precisely
because the finite-dimensional supremum need not be attained. Together
these imply that any general
structural criterion must be restricted to a subclass.

For the specific $I_{3322}$ equality module established here, a hypothetical
finite-dimensional maximizer would have a finite occupied response orbit;
the resulting closure caps its value at $1/4<S$. Extending this mechanism
to a dichotomy for other two-outcome bipartite functionals is a conjectural
direction, not a consequence of Theorem~\ref{thm:N}. Whether an appropriately
defined finite recurrence conversely
\emph{suffices} for finite-dimensional attainment we do not know; nor do we
have a definition of the critical transport for a general two-outcome
functional, and supplying one is the
first obstacle to testing it. We also
note the marginal/correlator boundary: Tsirelson's theorem covers
correlators only, and it is exactly the marginal terms through which
$I_{3322}$ escapes it.

Finally, two cautions from the literature, both
concerning the \emph{hierarchy} sense of ``attain'' rather than the
strategy sense used in this paper: non-saturation of finite levels of the
Navascu\'es--Pironio--Ac\'in hierarchy \cite{NPA} is not evidence of
strategy-nonattainment --- doubly-tilted CHSH functionals near the critical
line $\alpha+\beta=2$, where the boundary curvature degenerates, are attained
at $d=2$ yet are not saturated by NPA level~$10$, as reported in
\cite{Pakhunov} with the level-10 figure due to Gigena and Kaniewski
\cite{GigenaKaniewski} --- and the NPA hierarchy need not attain the commuting value
at any finite level \cite{Fanizza} (whose headline result is the
undecidability of deciding $\omega_{qc}>1/2$, in their notation for the
commuting-operator value).

Theorem~\ref{thm:rate} is the quantitative face of this picture: the value
carried by the open chain is approachable, but only at geometric cost,
$D(\eps)=\Theta(\log(1/\eps))$. For contrast, the embezzlement-based
witness of \cite{Coladangelo} --- a witness of the distinct separation
$\Cqs\neq C_{qa}$ --- requires local dimension
$2^{\Omega(\eps^{-1/8})}$; $I_{3322}$ therefore sits at the opposite,
logarithmic end of the known spectrum for the cost of approaching a value
that is never attained.

The exact identification of $S$ beyond the certified window remains open
here. Pauwels' Theorem~1 \cite{Pauwels} proves that the full optimized
finite P\'al--V\'ertesi family has supremum $S$; this is distinct from
identifying the limit of a prescribed numerical parameter sequence.
The amplitude-compatibility equation refuted in our correction history
belonged to this project's withdrawn construction. Its failure does not
refute the P\'al--V\'ertesi family. The spatial construction above uses an
independently extracted probability carrier.

\section{Methods and data}\label{sec:methods}

\begin{center}
\fbox{%
\begin{minipage}{0.94\textwidth}
\medskip
\begin{center}\textbf{\large Verifying this paper's claims}\end{center}
\raggedright
\begin{enumerate}\setlength{\itemsep}{3pt}\setlength{\parskip}{0pt}
\item \texttt{pip install -r requirements.txt}, then
  \texttt{python tools/check\_revision.py} --- development checks for
  this revision, not a final release certificate. The frozen-release
  verifier and its scope are documented in \texttt{VERIFY.md}.
\item Theorem~\ref{thm:N}:
  \texttt{certificate/production/theorem-N-four-receipts-at-S/} ---
  historical proof documents and exact-arithmetic guards. The current
  global-weld and occupied-component corrections are in
  Section~\ref{sec:N} and \texttt{paper/REVISION-NOTES.md}.
\item Theorem~\ref{thm:S}:
  \texttt{certificate/production/theorem-S-spatial-attainment-at-S/} ---
  proof documents and the amendment of 2026-08-07. The current
  parity-specific amplitude assignment is in Section~\ref{sec:S}.
\item Theorem~\ref{thm:rate}, the rate $D(\eps)=\Theta(\log(1/\eps))$:
  the new lower proof in Section~\ref{sec:rate-lower}, and the constructive
  upper proof in Section~\ref{sec:rate-upper}, with its source record in
  \texttt{certificate/production/rate-theta-log/upper-U1G-bundle/}.
\item Revised lower bound: Section~\ref{sec:rate-lower} contains the
  replacement proof; exact controls are in
  \texttt{certificate/production/lower-weighted-flow/}.
  The older lower bundle is preserved for history, not invoked for
  the quantitative conclusion.
\item Lean cores: \texttt{cd lean/I3322Kernel \&\& lake build} --- no
  \texttt{sorry}, standard axioms.
\item \texttt{VERIFY.md} at repository root: every claim mapped to its
  check and its analytic interfaces. \texttt{paper/CERTIFICATE-MAP.md}
  specifies correction precedence; \texttt{paper/BUILD.json} binds the
  current sources and PDF. The publication source manifest records
  the exact accompanying repository files.
\end{enumerate}
\medskip
\end{minipage}}
\end{center}

\subsection*{Exact computation}
The value-window certificates use rational arithmetic or outward-rounded
interval arithmetic. Historical independent-engine checks and current
replays have separate coverage: the current upper-window replay uses
the inspected production arithmetic, while the lower quotient is
reconstructed directly in rationals. The public verification guide records
the scope of these checks; a saved receipt is not a new replay.
Certificates and the correction record accompany the source repository
\cite{Repo}. The analytic steps are established in the proof documents; the guard
scripts check algebraic identities and exact arithmetic only. A Lean~4
kernel \cite{Lean} (\texttt{lean/I3322Kernel/}; with axiom reports in
\texttt{AxiomCheck.lean}, using only standard axioms) machine-checks
the quarter-ceiling chain of
Section~\ref{sec:N}, the band algebra of \S\ref{sec:rate-upper}, and the
three combinatorial cores associated with the historical lower-bound
route. The added \texttt{WeightedFlow} module checks selected finite
accounting lemmas used in \S\ref{sec:rate-lower}, including the all-edge
balance and the residual estimate conditional on a bounded potential.
Neither those lemmas nor the historical cores formalize the potential's
existence, the analytic estimates, or the complete quantum-to-flow argument.

\subsection*{Use of language models}
Large language models were used throughout this work --- for proof
discovery, implementation, verification engineering, adversarial auditing,
and editorial assistance --- across multiple model families and separate,
independently instantiated sessions, with the auditing sessions instructed to
refute rather than confirm. The author conceived the research program and its
hypotheses, set the verification standards and the preregistration and
correction protocols, made all promotion, retraction, and publication
decisions, and assumes sole responsibility for this work. The
proof is an analytic, model-assisted argument for which the author remains
responsible; exact computation does not by itself validate its analytic
interfaces. The numerical certificates use rational or outward-rounded
interval arithmetic, with the independently implemented checks documented
in the repository. The Lean kernel checks the selected
historical, upper-band, and finite weighted-flow lemmas specified above;
the complete replacement lower proof remains analytic, not a
Lean-certified theorem.

\subsection*{Data availability}
Repository: \url{https://github.com/Apsiape/i3322-exact-wall}. The historical
proofs and certificates are pinned to the archived release of record
(v4.0.0), version DOI
\href{https://doi.org/10.5281/zenodo.22099128}{10.5281/zenodo.22099128};
the concept
DOI \href{https://doi.org/10.5281/zenodo.21782008}{10.5281/zenodo.21782008}
resolves to the latest release and is not the object cited here.
The corrected proof is included in full in this paper and its accompanying
\texttt{LOWER-BOUND-REVISION.tex}. The accompanying repository's
\texttt{paper/REVISION-NOTES.md} and \texttt{VERIFY.md} give correction
precedence and reproduction instructions; \texttt{paper/BUILD.json}
identifies the sources used for its PDF. The older archive does not
contain this correction, and is cited for historical sources only.
The historical certificate directories, all under
\texttt{certificate/production/}, are
\begin{quote}\ttfamily\raggedright
theorem-N-four-receipts-at-S/\\
theorem-S-spatial-attainment-at-S/\\
rate-theta-log/
\end{quote}

\subsection*{Competing interests}
None.

\subsection*{Acknowledgments}
This work owes its problem to K.~F.~P\'al and T.~V\'ertesi, whose 2010
analysis defined the object and whose numerical family remains the guide to
its structure; a debt of direction to A.~Coladangelo and J.~Stark, whose
infinite-dimensional correlation and closing suggestion pointed at exactly
the theorem proved here; and it cites, as prior exact certification,
N.~Mghirbi's independently replayable $I_{3322}$ enclosures.

\end{document}